\documentclass{aa} 
\usepackage{graphicx}
\usepackage{txfonts}
\usepackage{natbib}
\usepackage{ulem}
\def\thco{$^{13}$CO}
\def\ceo{C$^{18}$O}
\def\hho{H$_2$O}

\def\hh{H$_2$}
\def\hhhp{H$_3^+$}
\def\ohp{OH$^+$}
\def\hhop{H$_2$O$^+$}
\def\hhhop{H$_3$O$^+$}

\def\mic{$\mu$m}
\def\kms{km\,s$^{-1}$}

\def\pow#1#2{#1$\times$10$^{#2}$}
\def\rs{s$^{-1}$}
\def\scm{cm$^{-2}$}
\def\ccm{cm$^{-3}$}

\def\lsol{L$_{\odot}$}
\def\lbol{L$_{\rm bol}$}
\def\dv{$\Delta${\it V}}

\def\vhel{$V_{\rm helio}$}

\def\txc{$T_{\rm ex}$}
\def\old#1{}
\def\new#1{#1}
\begin{document}
\title{The ionization rates of galactic nuclei and disks from Herschel/HIFI observations of water and its associated ions
\thanks{\textit{Herschel} is an ESA space observatory with science instruments provided
by European-led Principal Investigator consortia and with important participation from NASA}}
\titlerunning{The ionization rates of galactic nuclei and disks}

\author{F.F.S. van der Tak \inst{\ref{sron},\ref{rug}} \and
        A. Wei\ss \inst{\ref{mpifr}} \and
        L. Liu \inst{\ref{mpifr}} \and
        R. G\"usten \inst{\ref{mpifr}}
}

\institute{SRON Netherlands Institute for Space Research, Landleven 12, 9747 AD Groningen, The Netherlands; \email{vdtak@sron.nl} \label{sron} 
  \and Kapteyn Astronomical Institute, University of Groningen, The Netherlands \label{rug}
  \and Max-Planck-Institut f\"ur Radioastronomie, Auf dem H\"ugel 69, 53121 Bonn, Germany \label{mpifr} }

\date{Submitted 13 January 2016; revised 13 May 2016; accepted 15 June 2016}

\abstract
{Dense gas in galactic nuclei is known to feed central starbursts and AGN, but the properties of this gas are poorly known due to the high obscuration by dust.}
{Submm-wave spectroscopy of water and its associated ions is useful to trace the oxygen chemistry of the nuclear gas, in particular to constrain its ionization rate.}
{We present Herschel/HIFI spectra of the \hho\ 1113 GHz and \hhop\ 1115 GHz lines toward five nearby prototypical starburst/AGN systems, and \ohp\ 971 GHz spectra toward three of these. The beam size of 20$''$ corresponds to resolutions between 0.35 and 7 kpc.
}
{The observed line profiles range from pure absorption (NGC 4945, M82) to P~Cygni indicating outflow (NGC 253, Arp 220) and inverse P~Cygni indicating infall (Cen A). 
The similarity of the \hho, \ohp, and \hhop\ profiles to each other and to HI indicates that diffuse and dense gas phases are well mixed.
We estimate column densities assuming negligible excitation (for absorption features) and using a non-LTE model (for emission features), adopting calculated collision data for \hho\ and \ohp, and rough estimates for \hhop. 
Column densities range from $\sim$10$^{13}$ to $\sim$10$^{15}$ \scm\ for each species, and are similar between absorption and emission components, indicating that the nuclear region does not contribute much to the emission in these ground-state lines. 
The $N$(\hho)/$N$(\hhop) ratios of 1.4--5.6 indicate an origin of the lines in diffuse gas, and the $N$(\ohp)/$N$(\hhop) ratios of 1.6--3.1 indicate a low \hh\ fraction ($\approx$11\%) in the gas.
The low \hho\ abundance relative to \hh\ of $\sim$10$^{-9}$ may indicate enhanced photodissociation by UV from \old{the nuclei} \new{young stellar populations}, or freeze-out of \hho\ molecules onto dust grains.}
{We use our observations to estimate cosmic-ray ionization rates for our sample galaxies, adopting recent Galactic values for the average gas density and the ionization efficiency. 
We find $\zeta_{\rm CR}$ $\sim$3$\times$10$^{-16}$ \rs, similar to the value for the Galactic disk, but $\sim$10$\times$ below that of the Galactic Center and $\sim$100$\times$ below  estimates for AGN from excited-state \hhhop\ lines. 
We conclude that the ground-state lines of water and its associated ions probe primarily non-nuclear gas in the disks of these centrally active galaxies.
Our data thus provide evidence for a decrease in ionization rate by a factor of $\sim$10 from the nuclei to the disks of galaxies, as found before for the Milky Way.
}


\maketitle

\section{Introduction}
\label{s:intro}

The star formation rates in galactic nuclei and disks are regulated by the physical conditions in their interstellar media. 
In particular, the gas density sets the free-fall time for gravitational collapse, the kinetic temperature sets the (Jeans) mass scale for fragmentation of the collapsing cloud, while turbulence and magnetic fields may provide at least partial support against gravitational collapse.
See \citet{kennicutt2012} for a recent review.

The bulk of the star formation in galaxies takes place in dense interstellar clouds, and the determination of the conditions in such clouds require long-wavelength observations due to the high column densities of dust.
Recent advances in observing technology, in particular at submillimeter wavelengths, have led to rapid progress in the determination of gas temperatures and kinetic temperatures in galactic nuclei and disks.
Galaxies are now well known to exhibit significant differences between their nuclei and disks in the physical parameters of gas clouds, in the chemical composition of the gas, and in their star formation history (e.g., \citealt{gonzalez2012}).
However, the ionization rate of the gas in galaxies is poorly known, and variations of this rate within galaxies have not yet been explored, although the ionization fraction of gas clouds determines the dynamical importance of magnetic fields \citep{grenier2015}.
Furthermore, cosmic-ray ionization is a significant heating source for interstellar gas, and high ionization rates may lead to high gas temperatures and hence a top-heavy Initial Mass Function (IMF) as advocated by \citet{papadopoulos2010}.

Water and its associated ions (\ohp, \hhop, and \hhhop) are useful to trace the oxygen chemistry of interstellar gas clouds, and their ionization rates.
While optical lines of these ions are good ionization probes for diffuse lines of sight \citep{porras2014,zhao2015}, only their submillimeter transitions probe dense obscured regions.
Mapping of the SgrB2 region near the Galactic Center in the \hhhop\ 364 GHz line indicates an enhancement of the cosmic-ray ionization rate by an order of magnitude relative to the Solar neighbourhood value \citep{vdtak2006}. This enhancement is also seen in observations of \hhhp, which probe more diffuse gas \citep{oka2005}.
Observations of \hhhop\ toward active galactic nuclei (AGN) show even higher ionization rates, enhanced by another order of magnitude \citep{vdtak2008}, although with some uncertainty since the excitation of the line is not well constrained \citep{aalto2011}.
The advantage of \ohp\ and \hhop\ over \hhhop\ is that their lines lie at similar frequencies as \hho\ \old{so that the same gas volume is probed} \new{thereby minimizing beam size effects}, while using ground-state lines minimizes uncertainties through excitation effects.
However, although \ohp\ is observable from high dry sites such as Chajnantor \citep{wyrowski2010oh+}, the ground-state lines of \hho\ and \hhop\ are not accessible from the ground.

The Herschel mission \citep{pilbratt2010} offered the first opportunity to observe \hho\ and \hhop\ lines at high enough sensitivity to study external galaxies at high resolution. 
Observing the ground-state lines of \hho\ and \hhop\ requires high spectral resolution since their line profiles typically consist of a mixture of emission and absorption components (e.g., \citealt{benz2010,vdtak2013wish}).

This paper presents spectra of the \hho\ and \hhop\ ground-state lines near 1113 and 1115 GHz toward a sample of five nearby active galaxies (NGC 4945, NGC 253, Arp 220, M82, and Cen A) which may be considered as prototypical for their respective classes, and spectra of the \ohp\ ground-state line near 971 GHz toward three of these galaxies.
NGC 4945 is a dust-enshrouded Seyfert nucleus with a bolometric luminosity of \pow{2.4}{10} \lsol\ \new{(adopted from NED)} and an SED that peaks in the far-infrared.
NGC 253 is a starburst nucleus with \new{\lbol\ =}  \pow{1.7-2.1}{10} \lsol\ depending on the adopted distance (2.6--3.5 Mpc) and an SED that peaks in the optical.
Arp 220 is an ultraluminous ( \new{\lbol\ =} \pow{1.4}{12} \lsol) merger system with a double nucleus and high dust obscuration, causing the SED to peak in the far-infrared.
M82 is a starburst disk with  \new{\lbol\ =} \pow{5.3}{10} \lsol\ and an SED that peaks in the far-infrared.
We have observed three positions in M82: the center and 15$''$ offsets to the $\sim$NE and $\sim$SW (at PA = 72$^\circ$), which correspond to peaks in the CO emission.
Finally Cen~A (NGC 5128) is a radio AGN with  \new{\lbol\ =}  \pow{4.7}{11} \lsol\ and an SED which peaks in the X-ray band.

The outline of this paper is as follows: Section~\ref{s:obs} describes our observations, \S\ref{s:res} the observational results, \S\ref{s:disc} our derived physical parameters of the sources, and \S\ref{s:concl} our conclusions.

\section{Observations}
\label{s:obs}

\begin{table}
\caption{Source sample.}
\label{t:sample}
\begin{tabular}{cccccc}
\hline \hline
\noalign{\smallskip}
Source & $\alpha$(J2000) & $\delta$(J2000) & $d$  & ObsIDs \\
              & hh mm ss.ss         & $^0$ $'$ $''$         & Mpc & 1342... \\
\noalign{\smallskip}
\hline
\noalign{\smallskip}
NGC 253   & 00:47:33.12  & $-$25:17:17.6  & 3 & 212196, \\ 
                    &                         &                            &     & 210670 \\ 
M82c          & 09:55:52.22 &  $+$69:40:46.9 & 4  & 194792 \\ 
M82SW      & 09:55:49.44 & $+$69:40:42.3 &  4  & 203934 \\ 
M82NE       & 09:55:55.00 & $+$69:40:51.5 &  4  & 203933 \\ 
NGC 4945 & 13:05:27.48 &  $-$49:28:05.6 & 3.8 & 200995, \\ 
                    &                         &                            &       & 200946 \\ 
Cen A         & 13:25:27.61 &  $-$43:01:08.8 & 3.5 & 212190 \\ 
Arp 220      & 15:34:57.26  & $+$23:30:11.4 & 72 & 201000,\\ 
                    &                         &                             &      & 201001, \\
                    &                         &                             &       & 201559 \\  
\noalign{\smallskip}
\hline
\noalign{\smallskip}
\end{tabular}
\tablefoot{Distances are from the NASA Extragalactic Database (NED) at {\tt http://ned.ipac.caltech.edu/}.}
\end{table}

The positions in Table~\ref{t:sample} were observed in April-December 2010 using Band 4b of the HIFI instrument \citep{degraauw2010}, as part of the HEXGAL guaranteed time program (PI: G\"usten). 
We used the Double Beam Switch observing mode with a chopper throw of 3$'$.
The backend was the acousto-optical Wide-Band Spectrometer (WBS) which provides a bandwidth of 4$\times$1140 MHz (1200 \kms) at a resolution of 1.1 MHz (0.3 \kms).
This bandwidth is sufficient to simultaneously cover the p-\hho\ $1_{11}$--$0_{00}$ line at 1113.343 GHz (hereafter 1113 GHz) 
and the o-\hhop\ $1_{11}$--$0_{00}$ $J$=3/2-1/2 line at 1115.204 GHz (hereafter 1115 GHz).
The FWHM beam size at this frequency is 20$''$ \citep{roelfsema2012}, which corresponds to 7~kpc at the distance of Arp 220 and $\approx$0.34~kpc for the other sources.
The HIFI beam thus covers the Arp 220 system completely, while for the other galaxies, our observations only probe the nuclei and the disk gas in front of the nuclei.  
System temperatures ranged from 350 to 400 K and integration times from 120 to 536 minutes on-source.

Observations of the \ohp\ $N$=1--0 line at 971.804 GHz (hereafter 971 GHz) toward the nuclei of NGC 253, NGC 4945, and Arp 220 were also obtained within the HEXGAL program, using Band 4a of HIFI. 
For these data, system temperatures are 240--430 K and integration times are 12--38 minutes. 
The beam size of 22$''$ is very similar to that of the 1113-1115 GHz spectra, which permits a direct comparison of the results.

In addition to the data presented in \S\ref{s:res}, the \hho\ and \hhop\ lines were also unsuccessfully searched for in other galaxies, namely NGC 6240, NGC 4038/39, Mrk 231, NGC 1068, and M83. 
The ObsIDs of these data are 191679, 201002, 201067, 213331, and 213333.
Similarly, searches for the \hhop\ 625 GHz and \ohp\ 1892 GHz lines toward NGC 253 and NGC 4945 were unsuccessful.
The ObsIDs of these data are 210690, 200936, 201645, and 210791.

Calibration of the data was performed in the \textit{Herschel} Interactive Processing Environment (HIPE; \citealt{ott2010}) versions 10-12; further analysis was carried out in the CLASS\footnote{\tt http://www.iram.fr/IRAMFR/GILDAS} package, with the version of February 2013. 
Raw antenna temperatures were converted to $T_A$ scale using aperture efficiencies reported by \citet{roelfsema2012}, and linear baselines were subtracted. 
After inspection, the data from the two polarization channels were averaged to obtain rms noise levels of 3--7 mK per 5\,\kms\ channel for the \hho/\hhop\ setting, and 5--20 mK for the \ohp\ spectra. 
The absolute calibration uncertainty of HIFI Band 4 is estimated to be 10-15\%,
but the relative calibration between the \hho\ and \hhop\ lines in our spectra should be much better.

\section{Results}
\label{s:res}

\begin{figure}[tb]
\centering
\includegraphics[width=7cm,angle=0]{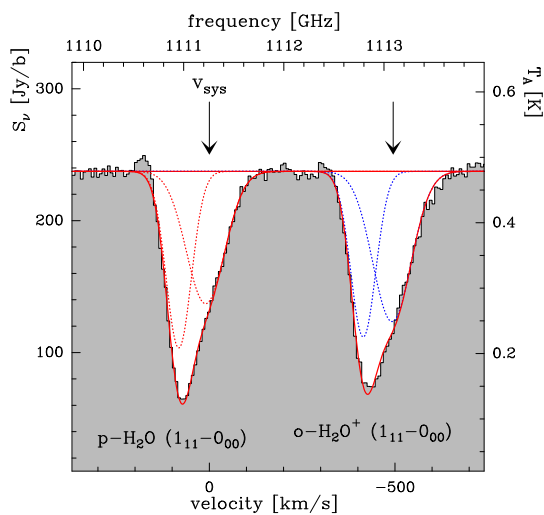}
\caption{Spectrum of the nucleus of NGC 4945 taken with Band 4b of Herschel-HIFI. Dotted lines indicate individual fit components and the solid line is the sum of these. The arrows indicate the systemic velocities for the \hho\ and \hhop\ lines. The velocity scale refers to the \hho\ line, and is relative to a systemic velocity of $V_0$=563\,\kms.}
\label{f:ngc4945}
\end{figure} 

\begin{figure}[tb]
\centering
\includegraphics[width=7cm,angle=0]{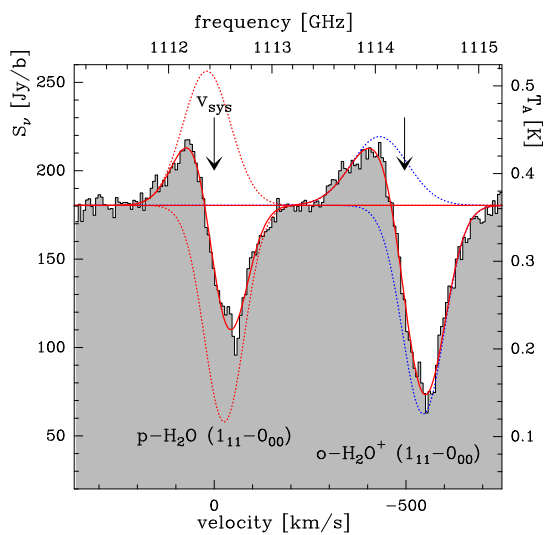}
\caption{As previous figure, for NGC 253, using $V_0$=243\,\kms.}
\label{f:ngc253}
\end{figure} 

\begin{figure}[tb]
\centering
\includegraphics[width=7cm,angle=0]{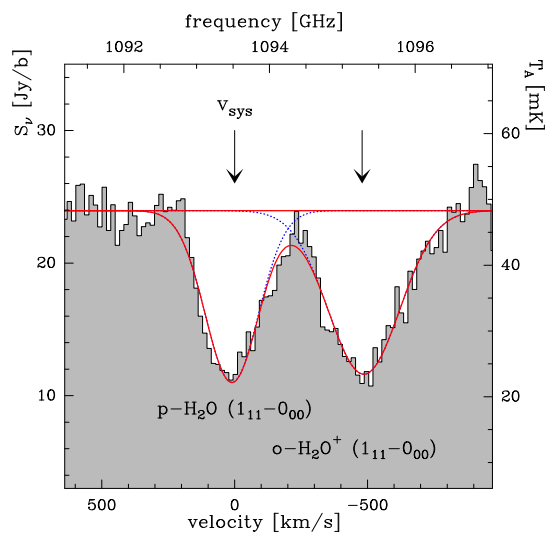}
\caption{As previous figure, for Arp 220, using $V_0$=5434\,\kms.}
\label{f:arp220}
\end{figure} 

\begin{figure}[tb]
\centering
\includegraphics[width=7cm,angle=0]{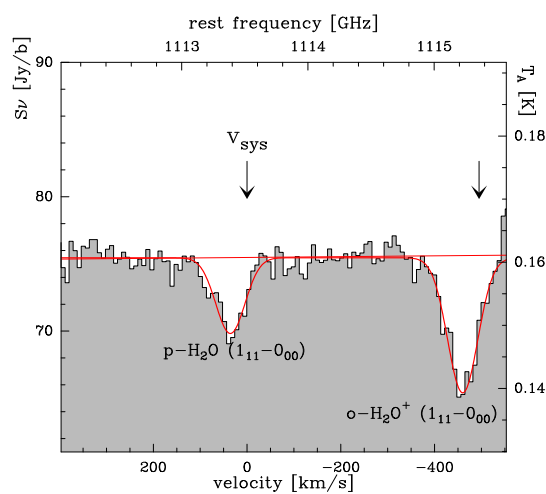}
\caption{As previous figure, for M82c, using $V_0$=203\,\kms.}
\label{f:m82c}
\end{figure} 

\begin{figure}[tb]
\centering
\includegraphics[width=7cm,angle=0]{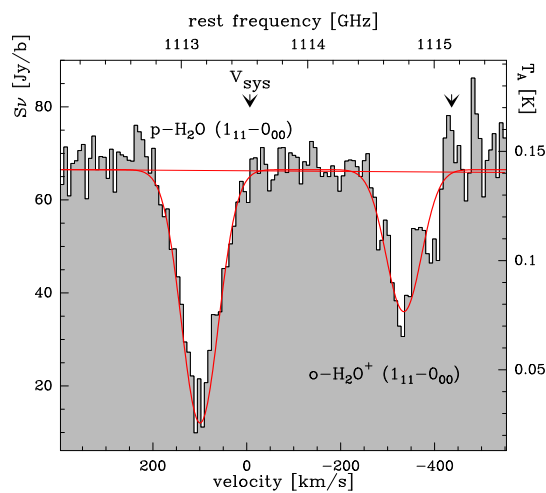}
\caption{As previous figure, for M82NE, using $V_0$=203\,\kms.}
\label{f:m82ne}
\end{figure} 

\begin{figure}[tb]
\centering
\includegraphics[width=7cm,angle=0]{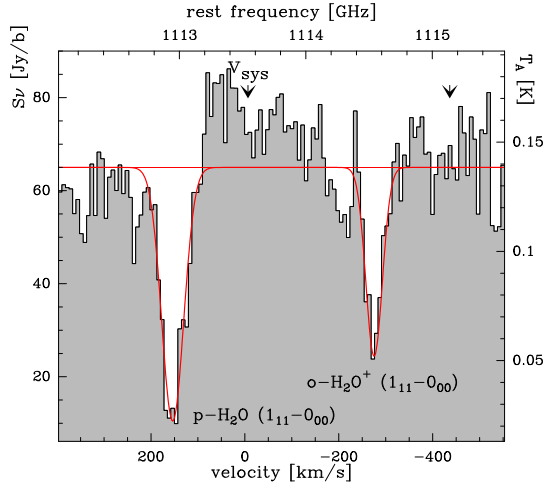}
\caption{As previous figure, for M82SW, using $V_0$=203\,\kms.}
\label{f:m82sw}
\end{figure} 

\begin{figure}[tb]
\centering
\includegraphics[width=7cm,angle=0]{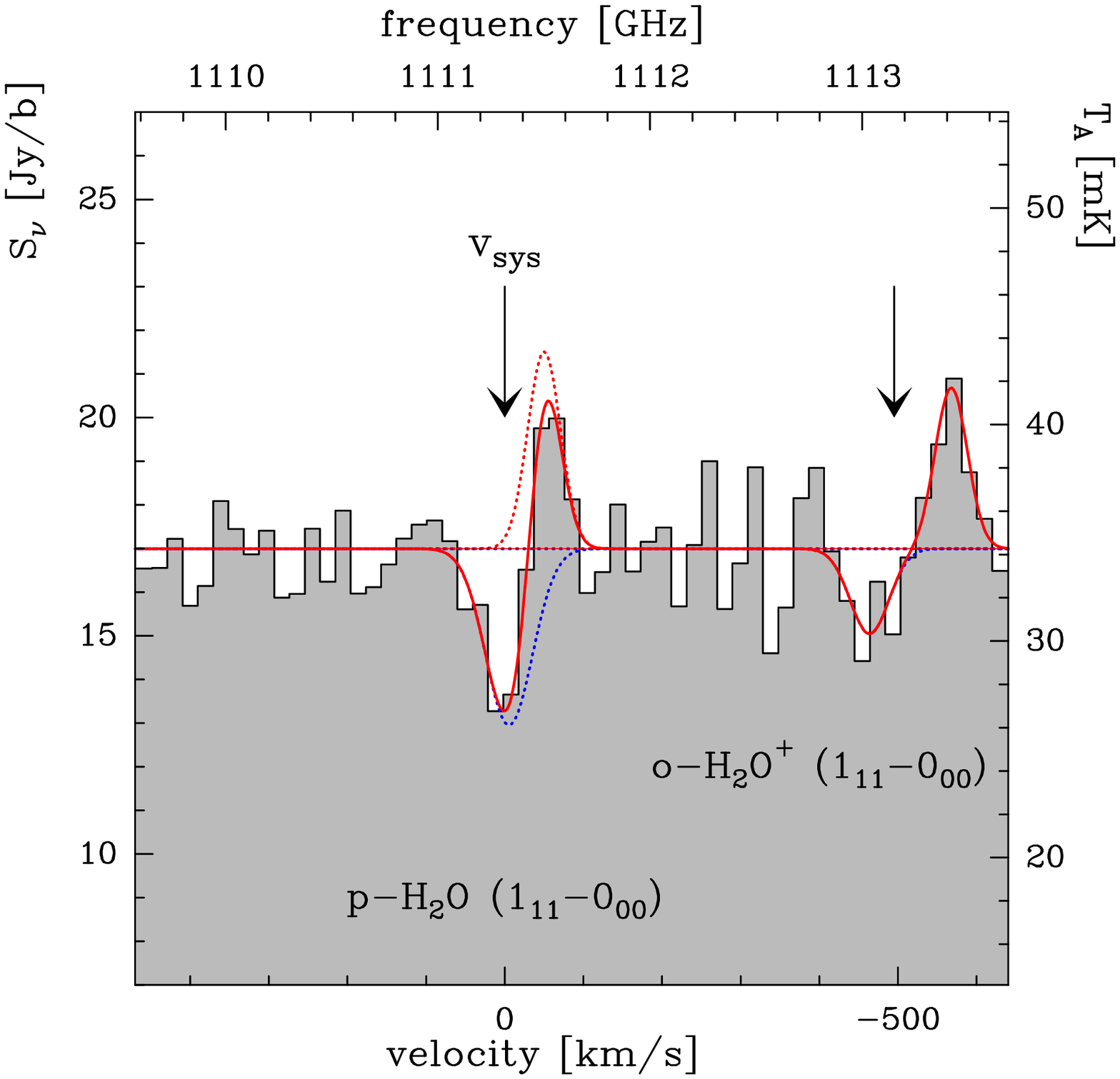}
\caption{As previous figure, for Cen A, $V_0$=547\,\kms.}
\label{f:cena}
\end{figure} 

\subsection{Line profiles}
\label{ss:profiles}

Figures~\ref{f:ngc4945}--\ref{f:cena} present the calibrated 1113-1115 GHz spectra, both in $S_\nu$ and $T_A$ units, on velocity and frequency axes. 
The \hho\ and \hhop\ lines are detected in all five sources, and strong continuum is also seen. 
The appearance of the lines differs from source to source as discussed below. 
The reported continuum temperature is half the observed value because continuum radiation enters the receiver through both sidebands while the line is only in one sideband.

Toward NGC 4945, the \hho\ and \hhop\ lines appear in absorption, the shape of which is well fitted with a double Gaussian. 
The derived parameters for the two lines are very similar to each other, and also similar to lines of HF observed with Herschel \citep{monje2014} and of HI observed with the VLA \citep{ott2001}. 
The broad absorption component is centered close to the systemic velocity, while the narrower component is redshifted by $\approx$80~\kms. 
As discussed by \citet{monje2014}, this appearance suggests an origin in a molecular gas ring that is possibly undergoing infall.
\new{Alternatively, redshifted absorption may result from non-circular gas motions, for instance in a barred potential.}

The spectrum of NGC 253 shows a combination of emission and absorption for both lines, although the central velocities of \hho\ and \hhop\ differ by 21--56\,\kms. 
Such P~Cygni profiles are indicative of gas outflow which is also seen in other species, perhaps most dramatically in OH with Herschel/PACS \citep{sturm2011}.
\new{Other species such as HF \citep{monje2014} and HI \citep{koribalski1995} show absorption profiles with a similar shape as the \hho\ and \hhop\ absorptions, but HI is without the corresponding emission feature.}

The ultraluminous merger Arp 220 shows a single broad absorption in both species, which is so broad that they almost overlap. The two lines probably trace the same gas although the \hhop\ line is somewhat broader than that of \hho. The \hhop\ profile may be influenced by a second absorption near --350 \kms, which is of marginal significance. The centroid position of both lines suggests an origin in the Western nucleus of the system \citep{aalto2007}, which is suspected to harbour a supermassive black hole \citep{downes2007}. Ground-based observations of \hhhop\ show a line at a similar velocity as \hho\ and \hhop, but with a smaller line width \citep{vdtak2008}.

The spectra toward the three positions in M82 show narrow absorption in the \hho\ and \hhop\ lines. 
The profiles toward M82c show a possible weak secondary absorption feature which is redshifted from the main absorption by $\approx$50 \kms. 
The \hho\ absorption toward M82NE appears partially filled in by emission, unlike the \hhop\ line and unlike the spectra at the other positions.
Toward M82SW, it is unclear whether the spectral excess between the \hho\ and \hhop\ lines is due to line emission or a baseline artifact; we do not discuss it further.

The absorptions toward the `offset' NE and SW positions are actually deeper than toward the `central' M82c position.
Comparison with the high-resolution CO 1--0 map by \citet{walter2002} shows that the NE and SW positions correspond to peaks in the CO distribution, while the central position is a local minimum in CO.
We conclude that the \hho\ and \hhop\ lines follow the distribution of molecular gas as traced by the CO 1--0 line.

The nearby radio galaxy Cen A appears to show inverse P~Cygni profiles in the \hho\ and \hhop\ lines, although the S/N ratio is limited. 
The derived parameters for the two lines are very similar, and the small velocity shift between the two lines is of marginal significance. 
The inverse P~Cygni profile is a sign of gas infall, which in the case of Cen A may well be toward the central AGN. 
The \hho\ and \hhop\ profiles are similar to those of CO and CI lines observed from the ground by \citet{israel2014}, except that the broad emission from the circumnuclear disk is not seen here. 
\old{According to} \new{Following the nomenclature of} Israel et al., the narrow \hho\ and \hhop\ emission likely originates in the 'extended thin disk' component of Cen A. 
The absorption components are also seen in CO \old{and CI}, but they are not discussed by Israel et al.

\begin{table}
\caption{Measured line parameters.}
\label{t:pars}
\begin{tabular}{llcccccc}
\hline \hline
\noalign{\smallskip}
Source & Line   & \vhel\  & \dv\     & $T_l$  & $T_c$\\
               &           & \kms\  & \kms\  & mK          & mK \\ 
\noalign{\smallskip}
\hline
\noalign{\smallskip}
NGC 4945 (n) & \hho\   & +645(1) & 80(1)  & --271(3) & 475 \\
NGC 4945 (n) & \hhop\ & +640(7) & 89(7)  & --278(26) & 475 \\
NGC 4945 (n) & \ohp\   & +654(5) & 70(5) &  --220(10) & 350 \\
NGC 4945 (b) & \hho\   & +573(1)  & 121(1)  & --203(1) & 475 \\
NGC 4945 (b) & \hhop\ & +557(7)  & 123(7)  & --196(14) & 475 \\
NGC 4945 (b) & \ohp\   & +574(5) & 120(5)  & --270(5) & 350 \\
NGC 253 (e) & \hho\   & +252(1) & 146(2) & +154(3) & 360 \\
NGC 253 (e) & \hhop\ & +308(6)  & 158(6) & +83(5) & 360 \\
NGC 253 (e) & \ohp\   & +301(24) & 176(18) & +130(4) & 230 \\
NGC 253 (a) & \hho\   & +217(1) & 122(1) & --248(2) & 360 \\
NGC 253 (a) & \hhop\ & +196(6)  & 132(6) & --234(12) & 360 \\
NGC 253 (a) & \ohp\   & +216(6)  & 158(5) & --290(9) & 230 \\
Arp 220 & \hho\   & +5445(5) & 235(11) & --26(2) & 48 \\
Arp 220 & \hhop\ & +5443(9) & 312(21) & --25(2) & 48 \\
Arp 220 & \ohp\   & +5418(4) & 283(9) & --25(1) & 33 \\
M82c     & \hho\   & +239(3) & 73(8) & --11(4) & 151 \\
M82c     & \hhop\ & +239(2) & 75(4) & --20(2) & 151 \\
M82SW & \hho\   & +209(3) & 41(7) & --18(4) & 120 \\
M82SW & \hhop\ & +217(3) & 57(5) & --24(3) & 120 \\
M82NE & \hho\   & +288(3) & 104(6) & --30(2) & 140 \\
M82NE & \hhop\ & +280(2) & 113(5) & --40(2) & 140 \\
Cen A (e) & \hho\   & +497(6) & 49(11) & +9(3) & 34 \\
Cen A (e) & \hhop\ & +477(6) & 48(14) & +8(3) & 34 \\
Cen A (a) & \hho\   & +542(9)    & 70(20) & --8(2) & 34 \\
Cen A (a) & \hhop\ & +580(15) & 58(19) & --4(1) & 34 \\
\noalign{\smallskip}
\hline
\noalign{\smallskip}
\end{tabular}
\tablefoot{Numbers in parentheses are formal errors from the Gaussian fits in units of the last decimal; (n) = narrow component, (b) = broad component, (e) = emission component, (a) = absorption component. The line temperatures in Column~5 are in $T_A$ units and are relative to the SSB continuum values reported in Column~6 which have a 10--15\% calibration uncertainty.}
\end{table}

\begin{figure}[t]
\centering
\includegraphics[width=6cm,angle=-90]{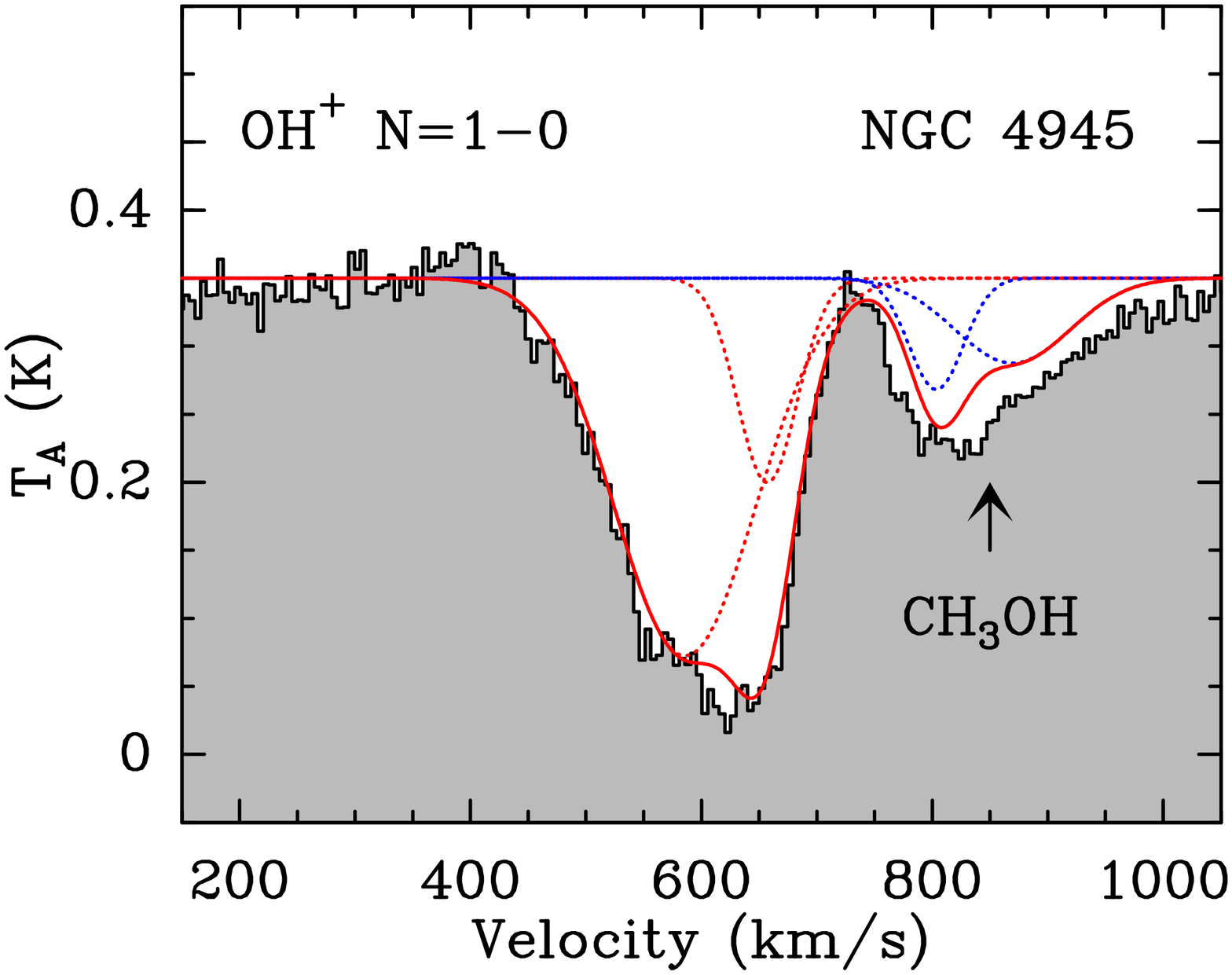}
\medskip
\includegraphics[width=6cm,angle=-90]{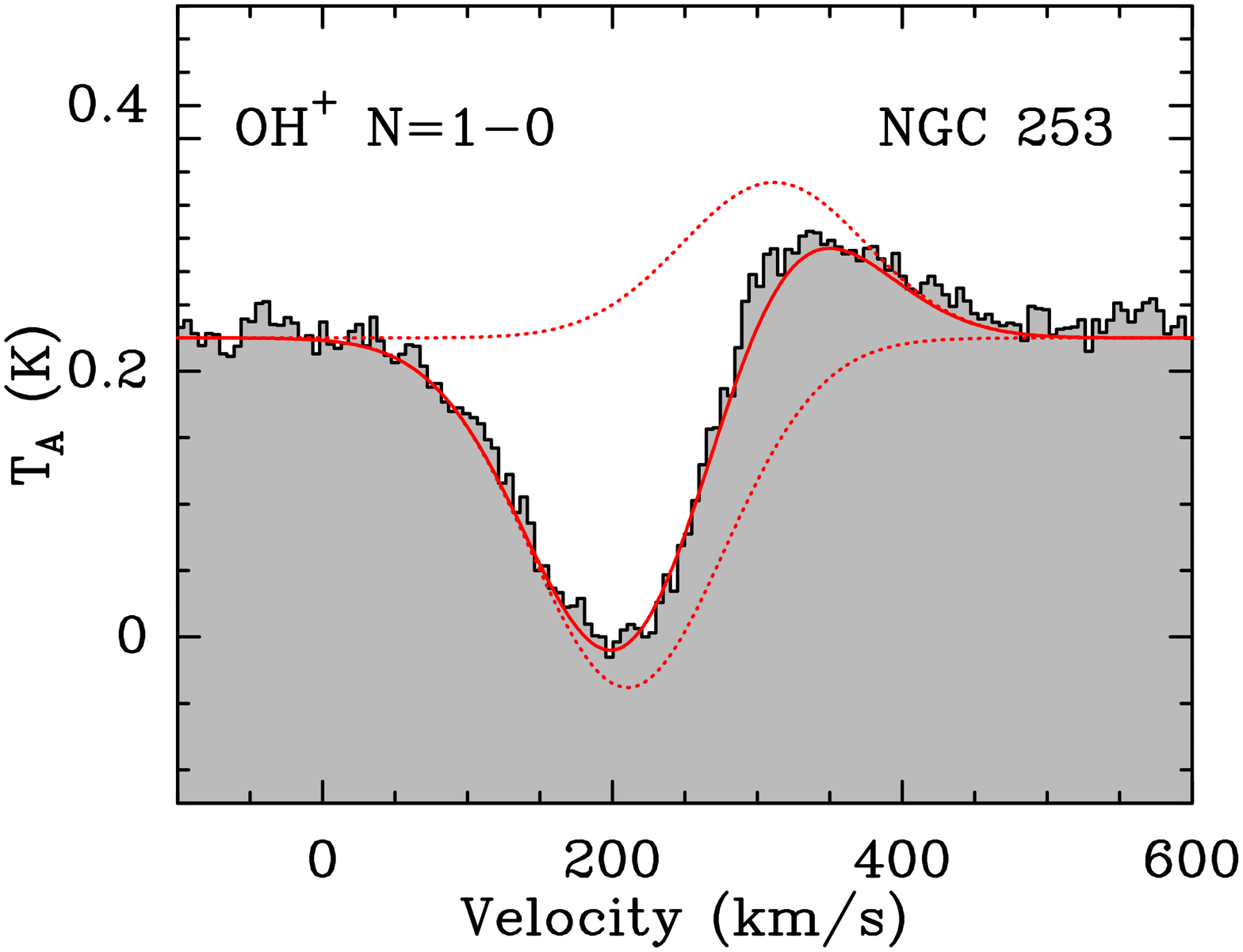}
\medskip
\includegraphics[width=6cm,angle=-90]{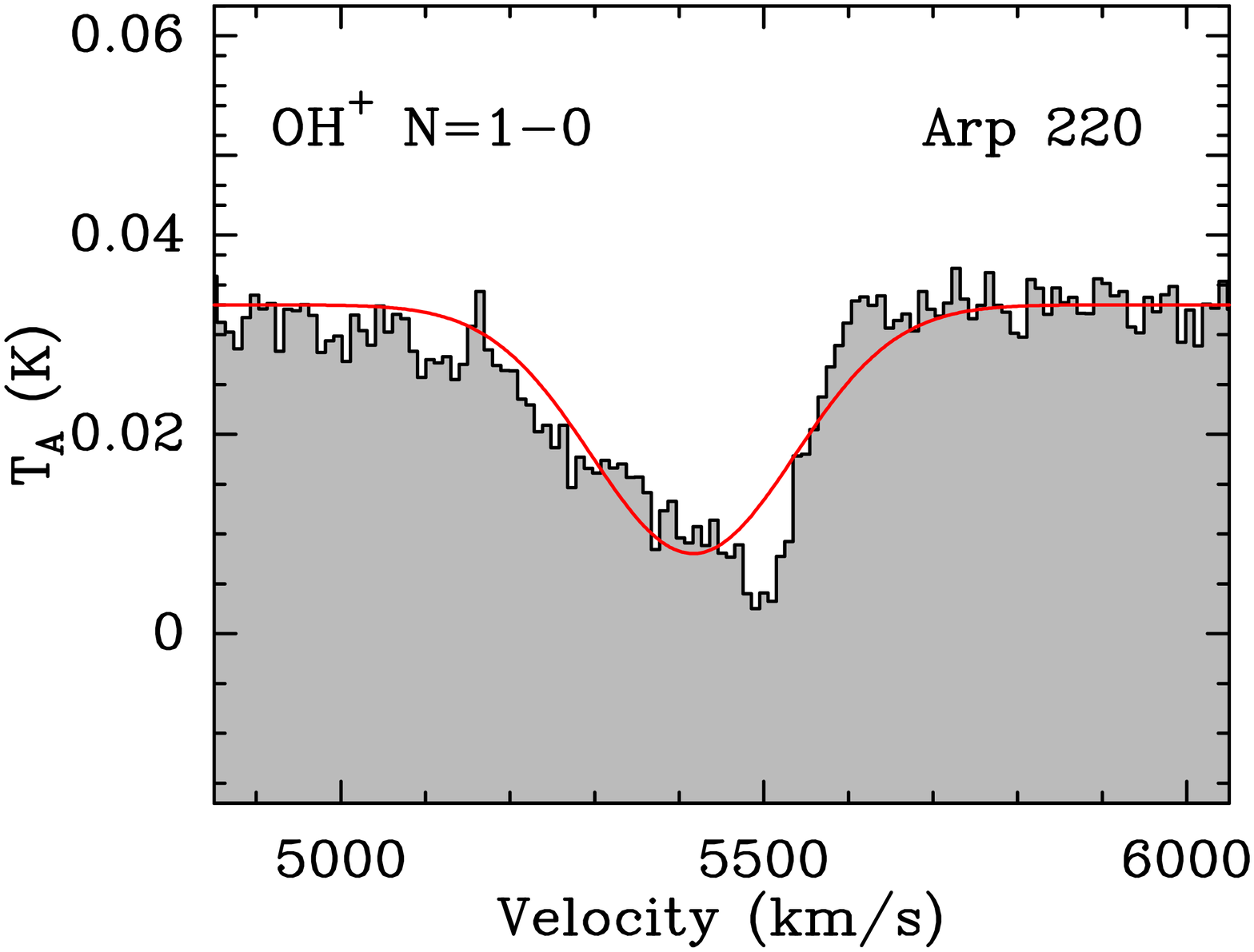}
\caption{Spectra of the \ohp\ 971 GHz line toward NGC 4945 (top), NGC 253 (middle), and Arp 220 (bottom).}
\label{f:oh+}
\end{figure} 

Figure~\ref{f:oh+} presents our spectra of the \ohp\ 971 GHz line toward NGC 4945, NGC 253, and Arp 220, where the line is clearly detected. 
The \ohp\ velocity profiles are similar to those of \hho\ and \hhop, with a double absorption toward NGC 4945 and a P Cygni-type line profile toward NGC 253.
The spectrum of Arp 220 shows a single broad absorption as for the \hho\ and \hhop\ lines, but the high signal-to-noise ratio reveals that this absorption has an asymmetric shape, which is indicative of an origin in a wind \new{or absorption by the second nucleus}.

The 971~GHz spectrum of NGC 4945 shows a second absorption feature right next to the \ohp\ line, which because of its mirrored shape probably originates from the image sideband.
The most likely candidate is the CH$_3$OH line at 959.4 GHz, since submm spectra of this galaxy show many lines from organic species \citep{wang2004}.
The NGC 4945 spectrum may also show a hint of an emission feature on the blueshifted shoulder of the \ohp\ line profile, which by itself is of marginal significance,
\new{but which may be real as it seems to have counterparts in \hho\ and possibly \hhop\ (Fig.~\ref{f:ngc4945}).}

\subsection{Column densities}
\label{ss:cold}

\begin{table*}
\caption{Beam-averaged column densities and abundances.}
\label{t:cold}
\begin{tabular}{lcccccc}
\hline \hline
\noalign{\smallskip}
Source & $N$(\hho)              & $N$(\hhop)             & \hho/\hhop\  & $N$(\ohp)               & \ohp/\hhop\  & $x$(\hho)  \\
              & $10^{14}$ \scm\   &  10$^{14}$ \scm\    &                       &  10$^{14}$ \scm\    &                       &   $10^{-9}$ \\ 
\noalign{\smallskip}
\hline
\noalign{\smallskip}
NGC 4945 (n) & 6.6 & 1.2 & 5.6 & 2.2 & 1.8 & 6.6 \\
NGC 4945 (b) & 6.6 & 1.6 & 4.2 & 3.2 & 2.0 & 6.6 \\
NGC 253 (a) & 14  & 2.7 & 5.2 & 8.4 & 3.1 & 9.3 \\
NGC 253 (e) & 8.8 & 1.6 & 5.5 & 0.9 & 0.5 & 5.9 \\
Arp 220 & 18 & 4.5 & 4.0 & 7.4 & 1.6 & 1.8 \\
M82c      & 0.56 & 0.21 & 2.6 & ... & ... & 0.62 \\
M82SW  & 0.66 & 0.13 & 5.3 & ... & ... & ... \\
M82NE  & 2.5 &   0.38  & 6.7 & ... & ... & ... \\
Cen A (a) & 0.19 & 0.14   & 1.4 & ... & ... & 1.1 \\
Cen A (e) & 0.18 & 0.043 & 4.2 & ... & ... & 1.1 \\
\noalign{\smallskip}
\hline
\noalign{\smallskip}
\end{tabular}
\tablefoot{\new{Uncertainties on column densities range from 10-15\% for absorption components (limited by calibration) to factors of 5--10 for emission components (through \txc).}}
\end{table*}

\new{Based on the appearance of the lines, we have fitted Gaussian models to the profiles, and Table~\ref{t:pars} presents the results. The asymmetric shape of the emission features suggests a geometry where some of the line emission is being absorbed by foreground material. To take this geometry into account, we simultaneously fit multiple Gaussians to the line profiles. The alternative of fitting separate Gaussians to the emission and absorption components would lead to lower line fluxes, but we consider this option less realistic.}

To estimate the column densities of the absorption components, we assume negligible excitation ($T_{\rm ex} < T_{10}$ where $T_{10} = E_{\rm up} / k_B \approx 50$\,K is the energy of the upper level above ground), so that only the molecular ground states are populated.
For \hho, this assumption is justified by the multi-line radiative transfer analysis by L.~Liu et al. (in prep.).
For \hho\ and \hhop, we assume an ortho/para ratio of 3 as observed in Galactic interstellar clouds by \citet{flagey2013} for \hho\ and \citet{schilke2013} for \hhop. 
Under these assumptions,
$$ N = N_l = \frac{g_l}{g_u} \frac{8\pi \tau^* \nu^3}{c^3 A_{ij}} $$
where $g_l$ and $g_u$ are the lower and upper state degeneracies, $\nu$ is the line frequency, and $A_{ij}$ is the Einstein~A coefficient.
These spectroscopic data are taken from the CDMS \citep{mueller2005}\footnote{\tt www.cdms.de} and JPL \citep{pickett1998}\footnote{\tt spec.jpl.nasa.gov} catalogs; 
for \ohp\ and \hhop, we sum the contributions of the individual hyperfine components which are blended in our data.
For a Gaussian profile, the velocity-integrated apparent optical depth is given by
$$\tau^* = \int \tau dV = -1.06 \Delta V \ln \left( \frac{T_c-T_l}{T_c} \right),$$ 
where the line and continuum antenna temperatures $T_l$ and $T_c$ \new{and the FWHM line width \dv} are reported in Table~\ref{t:pars}. 
This expression assumes that the absorbing material covers the background continuum source homogeneously, and hence provides a lower limit to the actual line opacity.
A patchy distribution with a higher opacity may be more realistic, as discussed by \citet{weiss2010} for the case of M82c.

To model the excitation of the emission components, we use the non-LTE radiative transfer program RADEX \citep{vdtak2007}\footnote{\tt www.personal.sron.nl/$\sim$vdtak/radex/index.shtml} which includes collisional and radiative excitation, and treats optical depth effects with an escape probability formalism.
For \hho\ we use state-of-the-art quantum-mechanically computed collision data with \hh\ from \citet{daniel2011} as reported on the LAMDA database \citep{schoeier2005}\footnote{\tt http://home.strw.leidenuniv.nl/$\sim$moldata/}. 
For \ohp\ we use collision rates with electrons from \citet{vdtak2013oh+} and use the rates with He from \citet{gomezcarrasco2014} to model inelastic collisions with H; collisions with \hh\ are mostly reactive in the case of \ohp.
For \hhop, detailed collisional calculations do not exist, so we approximate the collision rates of the \new{radiatively allowed} transitions as $Q_0 * S_{ij}$ where $Q_0$ is a characteristic downward rate coefficient  and $S_{ij}$ is the normalized radiative line strength out of upper level $i$ summed over all lower states, which enters the calculation of $A_{ij}$ from the microwave intensity.
For the \hhop-\hh\ system, we adopt $Q_0 \sim 10^{-10}$\,cm$^3$\,\rs\ since \hhop\ has a high dipole moment which should exhibit strong coupling to the \hh\ molecule.

We have run a grid of RADEX models covering kinetic temperatures from 10 to 100~K and \hh\ volume densities from $10^4$ to $10^6$\,\ccm; 
for \ohp, we assume $n$(H) = $n$(\hh) and $n$(e) / $n$(H) = 10$^{-4}$.
These ranges should bracket the likely physical conditions in the nuclei of our sample galaxies (see \S\ref{s:disc}).
For all three species, we find \txc\ values from 3.4~K at low temperature and density to $\approx$9~K at high temperature and density. 
The optical depths of the lines scale linearly with the molecular column density up to $N = 10^{14}$\,\scm, where $\tau=0.4$ is reached.
We conclude from these calculations that under the likely conditions in the gas studied here, we may expect low excitation temperatures (5--10~K) for our studied species, which is indeed well below $T_{10}$ as assumed above for the absorption components.
A second conclusion is that for these lines, the optically thin assumption is only valid for column densities up to $\sim$10$^{14}$\,\scm.

As an alternative to collisions, radiative pumping may dominate the excitation of the emission components, as found for high-$J$ lines of \hho\ in data from SPIRE \citep{yang2013}, PACS \citep{gonzalez2014}, and HIFI (L. Liu et al., in prep.). 
\old{In the following}
\new{For our 3 species}, we model this case by adopting \txc\ = 100\,K for all lines, which lowers our estimates of the total column density by factors of $\approx$15 relative to the values for \txc\ = 10\,K. 
Further increasing \txc\ does not change the estimates much, as we approach the limit \txc\ $>>$ $T_{10}$.

\subsection{Abundances}
\label{ss:abun}

Table~\ref{t:cold} lists our derived column densities, where we adopt \txc\ = 10 K for emission components, as found above \new{as an average value between the cases of collisional and radiative} excitation, and use the formula from \citet{vdtak2013oh+}.
\new{The uncertainties on the column densities are factors of 5--10 for emission components, mainly through the uncertain \txc. 
The absorption column densities have small uncertainties, 10--15\%, limited by calibration error, if the absorbers fill most of the HIFI beam, but the uncertainty is higher if the covering fraction is small and the absorption saturated. 
Uncertainties on column density ratios should be much smaller than for absolute column densities, because the various sources of error cancel out at least partially.}

The column density values range over a factor of $\sim$100 for the species that are detected in all sources: from \pow{4}{12} to \pow{4.5}{14} \scm\ for \hhop,  and from \pow{1.8}{13} to \pow{1.8}{15} \scm\ for \hho. 
The column density variation is smaller for \ohp, which is only seen in three of our sources.
Our \hho\ and \hhop\ column densities for M82c are 5--10$\times$ lower than those in \citet{weiss2010} because we use the observed continuum instead of a model value.
The column densities toward the M82SW position are within 30\% of those toward M82c, while those toward M82NE are $\sim$2--4$\times$ higher, which follows the trend in the CO 1--0 emission map by \citet{walter2002}.
In the cases where line emission is detected, the derived column densities are $\sim$10--90\% of the value estimated from the corresponding absorption feature, with an average value of 52\%. 
The case of \ohp\ in NGC 253 follows this trend since the absorption feature is saturated so its column density is a lower limit.
The fact that similar columns of gas (to a factor of 2) are probed in emission and absorption suggests that the nuclear region does not contribute much to the line emission, at least in the ground-state transitions.

Table~\ref{t:cold} also presents column density ratios of \hho\ and \ohp\ over \hhop. 
The \hho/\hhop\ ratios range from 1.4 to 5.6, which is at the low end of the distribution of \hho/\hhop\ ratios for Galactic sources compiled by \citet{wyrowski2010h2o+}. 
Studying lines of sight toward high-mass star-forming regions, these authors found \hho/\hhop\ ratios of $\sim$3 toward low-density gas (foreground clouds and outflow lobes) and higher ratios ($\sim$10) toward dense protostellar envelopes.
Comparing our results with this Galactic study, our low observed \hho/\hhop\ ratios suggest an origin of the emission and absorption features in diffuse gas clouds.

The right-hand column of Table~\ref{t:cold} presents estimates of the \hho\ abundances toward our sources. 
These were obtained by dividing the $N$(\hho) values by estimates of $N$(\hh) in similar-sized telescope beams. 
For NGC 4945, we adopt $N$(\hh) = \pow{1.0}{23} \scm, based on \thco\ and \ceo\ observations at 230 GHz with the 15-m SEST telescope \citep{chou2007}.
For NGC 253, \citet{weiss2008} estimate $N$(\hh) = \pow{1.5}{23} \scm\ based on APEX/LABOCA maps of 870~\mic\ dust emission.
For Arp~220, we use $N$(\hh) = \pow{1}{24} \scm\ as derived by \citet{downes2007} for the extended disk/torus surrounding the Western nucleus (cf.\ \citealt{aalto2009}).
For M82c, we adopt $N$(\hh) = \pow{9}{22} \scm\ based on the discussion in \citet{weiss2010}.
For Cen~A, \citet{israel2014} estimate $N$(\hh) = \pow{1.7}{22} \scm\ for the circumnuclear disk with a 20$''$ diameter.
The resulting abundances may be considered lower limits, as they compare the total amount of \hh\ to the fraction of gas in absorption (the foreground).
However, the fact that we find similar column densities in emission and absorption suggests that this effect is small.

The \hho\ abundances in Table~\ref{t:cold} range from \pow{6.2}{-10} to \pow{9.3}{-9} which is $\sim$5--80$\times$ lower than typical values for Galactic diffuse clouds, where the \hho\ abundance is limited by photodissociation \citep{flagey2013}.
Our abundance values are in the normal range for cold dense Galactic gas clouds, where \hho\ is depleted on the surfaces of dust grains \citep{vandishoeck2013},
and somewhat below the usual values for warm dense gas (10$^{-8}$--10$^{-6}$: e.g., \citealt{emprechtinger2013}). 
However, an origin of the \hho\ lines in dense clouds appears inconsistent with our observed \hho/\hhop\ ratios, which are well below the value of $\sim$10 for dense gas. 
One possibility may be that the \hho\ lines arise in dense clumps embedded in more diffuse gas where the \ohp\ and \hhop\ lines originate, as suggested earlier to explain observed CO/CI ratios (e.g., \citealt{kramer2005}).
The similar shapes of the \ohp\ and \hhop\ velocity profiles to those of both \hho\ and HI then indicate that the diffuse and dense gas phases are well mixed.
We conclude that while the \ohp\ and \hhop\ lines likely arise in diffuse gas clouds, the origin of the \hho\ lines is somewhat uncertain.
 
\section{Discussion}
\label{s:disc}

To estimate the cosmic-ray ionization rates in our sample of galaxies from our \hhop\ and \ohp\ observations, we assume steady-state ion-molecule chemistry (e.g., \citealt{hollenbach2012}).
Table~\ref{t:chem} lists the relevant reactions and our adopted rates, which are taken from the UMIST database for astrochemistry \citep{mcelroy2013}\footnote{\tt http://www.udfa.net.}.
In diffuse interstellar clouds, cosmic-ray ionization of H leads to H$^+$, which charge transfers with O to form O$^+$, which reacts with \hh\ to form \ohp. The \ohp\ ion may recombine with a free electron (reaction 2) or react with \hh\ into \hhop\ (reaction 1), which may itself recombine with an electron (reaction 4) or react with \hh\ into \hhhop\ (reaction 3), which recombines into \hho\ and other products.

As shown by \citet{indriolo2012}, the cosmic-ray ionization rate may be written in terms of observables as 
$$ \epsilon \zeta_H = \frac{N({\rm OH}^+)}{N({\rm H})} n_H \left( \frac{f({\rm H}_2)}{2}k_1 + x_e k_2 \right),$$
where the molecular hydrogen fraction $f$(\hh), defined as
$$ f({\rm H}_2) = \frac{2n({\rm H}_2)}{n({\rm H}) + 2n({\rm H}_2)},$$ 
is related to the \ohp/\hhop\ ratio by
$$ \frac{N({\rm OH}^+)}{N({\rm H}_2{\rm O}^+)} = \frac{k_3 + 2k_4x_e / f({\rm H}_2)}{k_1}.$$

In these equations, $k_i$ is the rate of reaction $i$ as listed in Table~\ref{t:chem}, \old{evaluated at $T$=100~K,} and $\epsilon$ is the \ohp\ formation efficiency, which accounts for backward charge transfer from O$^+$ to H and the neutralization of H$^+$ on PAHs and dust grains \citep{neufeld2010}.
The only observational determination of this efficiency, based on \ohp, \hhop\ and \hhhp\ spectroscopy toward the W51 region \citep{indriolo2012}, indicates $\epsilon \approx 7$\%, which is in the range predicted by recent chemical models \citep{hollenbach2012}.
\new{The reaction rates in Table~\ref{t:chem} have been evaluated at an assumed typical temperature of $T$=100~K, but our results are insensitive to this assumption as the reactions with \hh\ ($k_1$ and $k_3$) are independent of $T$ and the recombination rates ($k_2$ and $k_4$) have a weak $T^{-0.5}$ dependence \citep{mcelroy2013}.}

By the last equation, our observed \ohp/\hhop\ column density ratios of 1.6--3.1 (Table~\ref{t:cold}, column 6) indicate molecular fractions from $f$(\hh) = 8\% for the absorption component of NGC 253 to 20\% for Arp 220. 
\old{For galaxies where \ohp\ data are unavailable, we assume an}
\new{Our average \ohp/\hhop\ ratio of 2.5} corresponds to $f$(\hh) = 11\%, which is somewhat higher than the typical value of 4\% for Galactic clouds \citep{indriolo2015}.
This result depends only on the column density ratio, so that any underestimates of the absolute column densities (e.g., due to filling factor assumptions) would cancel out.
For the diffuse type of interstellar clouds where the \ohp\ and \hhop\ lines thus likely originate, the electron abundance $x_e$ should equal the abundance of carbon since essentially all electrons come from photoionization of C. 
We adopt $x_e$ = \pow{1.5}{-4} \citep{sofia2004} although values up to \pow{2}{-4} may be possible \citep{sofia2011}.

The final parameter needed to estimate the cosmic-ray ionization rate is the column density of atomic hydrogen.
Although [HI] 21\,cm observations exist for all our galaxies, high-resolution data in beams matching our HIFI observations only exist for M82 \citep{yun1993} and Cen~A \citep{vdhulst1983}. 
These data indicate $N$(H) values that are 3--5 times lower than the corresponding $N$(\hh) numbers, suggesting that integrated over the line of sight, most gas is dense.
On the other hand, the low $f$(\hh) values derived above indicate that locally in the \ohp- and \hhop-absorbing clouds, most gas is diffuse.
The similarity of the \ohp\ and \hhop\ line profiles to those of \hho\ as well as those of HI (\S\,\ref{ss:profiles}) further indicates that the dense and diffuse gas phases are well mixed.
Taken together, the data suggest a picture where the \ohp-\hhop\ absorbers are diffuse pockets or 'bubbles' in a sea of dense gas.
\old{For these absorbers, we adopt $N$(H)/$N$(\hh) = 1/4 for galaxies where matched-beam HI data are unavailable, and} 
Assuming a typical gas density of $n_H$ = 35 \ccm\ \citep{indriolo2015},
\old{With these model parameters,} 
the column densities in Table~\ref{t:cold} indicate cosmic-ray ionization rates between \pow{6}{-17} and \pow{8}{-16} \rs, with an average value of \pow{3.4}{-16} \rs.
\new{The uncertainty on these values is substantial (factors of 2-3), both through the absolute column densities (Table~\ref{t:cold}) and the model assumptions. 
For galaxies where \ohp\ or matched-beam HI data are unavailable, the uncertainty is even higher (factors of 3-5), as average \ohp/\hhop\ or H/\hh\ column density ratios must be assumed.
These cases do not bias our results toward high or low ionization rates, so that our average value seems to be representative for our sample of sources.}

\begin{table}
\caption{Chemical network.}
\label{t:chem}
\begin{tabular}{rcll}
\hline \hline
\noalign{\smallskip}
\multicolumn{3}{c}{Reaction} & Rate (cm$^3$\rs)  \\
\noalign{\smallskip}
\hline
\noalign{\smallskip}
\ohp\ + \hh\ & $\to$ & \hhop\ + H & $k_1$ = \pow{1.01}{-9} \\ 
\ohp\ + e$^-$ & $\to$ & H + O & $k_2$ = \pow{1.1}{-8} \\
\hhop\ + \hh\ & $\to$ & \hhhop\ + H & $k_3$ = \pow{6.4}{-10} \\
\hhop\ + e$^-$ & $\to$ & products$^a$ & $k_4$ = \pow{6.8}{-7} \\
\noalign{\smallskip}
\hline
\noalign{\smallskip}
\end{tabular}
\tablefoot{$^a$: \hhop\ may recombine into H + OH or H + O + H; since the 'branching ratio' between these channels is irrelevant here, we just list the total reaction rate.}
\end{table}

\section{Conclusions}
\label{s:concl}

The cosmic-ray ionization rates derived in \S\ref{s:disc} are $\sim$100$\times$ below estimates for molecular gas in AGN from excited-state \hhhop\ emission in the far-infrared \citep{gonzalez2013}, which themselves are in line with the high $\gamma$-ray fluxes, radio synchrotron luminosities, and supernova rates in such systems (e.g., \citealt{persic2012}).
Our $\zeta_H$ values are also $\sim$10$\times$ below estimates for the Galactic Center from \hhhp\ mid-infrared absorption \citep{goto2014}, from \hhhop\ submm emission \citep{vdtak2008}, and from \ohp\ and \hhop\ far-infrared absorption \citep{indriolo2015}.
They are, however, similar to values derived for the disk of our Galaxy from \hhhp\ absorption lines \citep{mccall2012}, from HCO$^{+}$ submm emission \citep{vdtak2000} and from \ohp\ and \hhop\ absorption \citep{indriolo2015}.

Together with the apparent low excitation state of the molecules (\S\ref{ss:cold}), these low inferred $\zeta_H$ values suggest that the low-$J$ lines of \ohp, \hhop, and \hho\ trace the extended gas in the disks of our galaxies, rather than the warm dense circumnuclear gas probed by their high-excitation lines.
The gas probed by the low-$J$ lines may either be physically distant from any nuclear activity, or it may be shielded from its radiation by large columns of dust, as also seen in the Spitzer mid-infrared spectra of obscured AGN \citep{lahuis2007}.
The observed (inverse) P~Cygni profiles make the second option appear the most likely.

We conclude that the cosmic-ray ionization rate is not one constant that applies for a galaxy as a whole. 
Instead, $\zeta_H$ appears to vary by a factor of $\sim$10 between the disk of a galaxy and its nucleus; variations of similar magnitude appear to exist between galaxies. 
These variations are in line with changes in chemical composition between the disks and the nuclei of galaxies (e.g., \citealt{gonzalez2012}) and also with the different $\zeta_H$ values found for the disk of our Galaxy and its nucleus \citep{vdtak2006, indriolo2015}.

In the near future, ALMA and JWST will be instrumental to explore further differentiation between the various components of galaxies in terms of their physical and chemical conditions, including their ionization rates.
In the longer term, the SPICA mission will extend the present work to systems at higher redshift, and characterize the interstellar media of galaxies out to the peak of cosmic star formation at $z$=2.

\begin{acknowledgements}

The authors thank Russ Shipman (SRON) for help with data reduction, John Black (Onsala) and Fred Lahuis (SRON) for useful discussions, Fran\c{c}ois Lique (Le Havre) for sending \ohp-He collision data, and \new{Paul van der Werf \& Harold Linnartz (Leiden)} for comments on the manuscript.
This research has used the following databases: NED, SIMBAD, ADS, CDMS, JPL, and LAMDA.

\\

\par HIFI was designed and built by a consortium of institutes and university departments from across Europe, Canada and the US under the leadership of SRON Netherlands Institute for Space Research, Groningen, The Netherlands with major contributions from Germany, France and the US. Consortium members are: Canada: CSA, U.Waterloo; France: CESR, LAB, LERMA, IRAM; Germany: KOSMA, MPIfR, MPS; Ireland, NUI Maynooth; Italy: ASI, IFSI-INAF, Arcetri-INAF; Netherlands: SRON, TUD; Poland: CAMK, CBK; Spain: Observatorio Astron\'omico Nacional (IGN), Centro de Astrobiolog\'{\i}a (CSIC-INTA); Sweden: Chalmers University of Technology - MC2, RSS \& GARD, Onsala Space Observatory, Swedish National Space Board, Stockholm University - Stockholm Observatory; Switzerland: ETH Z\"urich, FHNW; USA: Caltech, JPL, NHSC.
\end{acknowledgements}

\bibliographystyle{aa}
\bibliography{hexgal}

\begin{thebibliography}{60}
\expandafter\ifx\csname natexlab\endcsname\relax\def\natexlab#1{#1}\fi

\bibitem[{{Aalto} {et~al.}(2011){Aalto}, {Costagliola}, {van der Tak}, \&
  {Meijerink}}]{aalto2011}
{Aalto}, S., {Costagliola}, F., {van der Tak}, F., \& {Meijerink}, R. 2011,
  \aap, 527, A69

\bibitem[{{Aalto} {et~al.}(2007){Aalto}, {Spaans}, {Wiedner}, \&
  {H{\"u}ttemeister}}]{aalto2007}
{Aalto}, S., {Spaans}, M., {Wiedner}, M.~C., \& {H{\"u}ttemeister}, S. 2007,
  \aap, 464, 193

\bibitem[{{Aalto} {et~al.}(2009){Aalto}, {Wilner}, {Spaans}, {Wiedner},
  {Sakamoto}, {Black}, \& {Caldas}}]{aalto2009}
{Aalto}, S., {Wilner}, D., {Spaans}, M., {et~al.} 2009, \aap, 493, 481

\bibitem[{{Benz} {et~al.}(2010){Benz}, {Bruderer}, {van Dishoeck},
  {St{\"a}uber}, {Wampfler}, {Melchior}, {Dedes}, {Wyrowski}, {Doty}, {van der
  Tak}, {B{\"a}chtold}, {Csillaghy}, {Megej}, {Monstein}, {Soldati},
  {Bachiller}, {Baudry}, {Benedettini}, {Bergin}, {Bjerkeli}, {Blake},
  {Bontemps}, {Braine}, {Caselli}, {Cernicharo}, {Codella}, {Daniel}, {di
  Giorgio}, {Dieleman}, {Dominik}, {Encrenaz}, {Fich}, {Fuente}, {Giannini},
  {Goicoechea}, {de Graauw}, {Helmich}, {Herczeg}, {Herpin}, {Hogerheijde},
  {Jacq}, {Jellema}, {Johnstone}, {J{\o}rgensen}, {Kristensen}, {Larsson},
  {Lis}, {Liseau}, {Marseille}, {McCoey}, {Melnick}, {Neufeld}, {Nisini},
  {Olberg}, {Ossenkopf}, {Parise}, {Pearson}, {Plume}, {Risacher},
  {Santiago-Garc{\'{\i}}a}, {Saraceno}, {Schieder}, {Shipman}, {Stutzki},
  {Tafalla}, {Tielens}, {van Kempen}, {Visser}, \& {Y{\i}ld{\i}z}}]{benz2010}
{Benz}, A.~O., {Bruderer}, S., {van Dishoeck}, E.~F., {et~al.} 2010, \aap, 521,
  L35

\bibitem[{{Chou} {et~al.}(2007){Chou}, {Peck}, {Lim}, {Matsushita}, {Muller},
  {Sawada-Satoh}, {Dinh-V-Trung}, {Boone}, \& {Henkel}}]{chou2007}
{Chou}, R.~C.~Y., {Peck}, A.~B., {Lim}, J., {et~al.} 2007, \apj, 670, 116

\bibitem[{{Daniel} {et~al.}(2011){Daniel}, {Dubernet}, \&
  {Grosjean}}]{daniel2011}
{Daniel}, F., {Dubernet}, M.-L., \& {Grosjean}, A. 2011, \aap, 536, A76

\bibitem[{{De Graauw} {et~al.}(2010){De Graauw}, {Helmich}, {Phillips},
  {Stutzki}, {Caux}, {Whyborn}, {Dieleman}, {Roelfsema}, {Aarts}, {Assendorp},
  {Bachiller}, {Baechtold}, {Barcia}, {Beintema}, {Belitsky}, {Benz}, {Bieber},
  {Boogert}, {Borys}, {Bumble}, {Ca{\"i}s}, {Caris}, {Cerulli-Irelli},
  {Chattopadhyay}, {Cherednichenko}, {Ciechanowicz}, {Coeur-Joly}, {Comito},
  {Cros}, {de Jonge}, {de Lange}, {Delforges}, {Delorme}, {den Boggende},
  {Desbat}, {Diez-Gonz{\'a}lez}, {di Giorgio}, {Dubbeldam}, {Edwards},
  {Eggens}, {Erickson}, {Evers}, {Fich}, {Finn}, {Franke}, {Gaier}, {Gal},
  {Gao}, {Gallego}, {Gauffre}, {Gill}, {Glenz}, {Golstein}, {Goulooze},
  {Gunsing}, {G{\"u}sten}, {Hartogh}, {Hatch}, {Higgins}, {Honingh}, {Huisman},
  {Jackson}, {Jacobs}, {Jacobs}, {Jarchow}, {Javadi}, {Jellema}, {Justen},
  {Karpov}, {Kasemann}, {Kawamura}, {Keizer}, {Kester}, {Klapwijk}, {Klein},
  {Kollberg}, {Kooi}, {Kooiman}, {Kopf}, {Krause}, {Krieg}, {Kramer},
  {Kruizenga}, {Kuhn}, {Laauwen}, {Lai}, {Larsson}, {Leduc}, {Leinz}, {Lin},
  {Liseau}, {Liu}, {Loose}, {L{\'o}pez-Fernandez}, {Lord}, {Luinge}, {Marston},
  {Mart{\'{\i}}n-Pintado}, {Maestrini}, {Maiwald}, {McCoey}, {Mehdi}, {Megej},
  {Melchior}, {Meinsma}, {Merkel}, {Michalska}, {Monstein}, {Moratschke},
  {Morris}, {Muller}, {Murphy}, {Naber}, {Natale}, {Nowosielski}, {Nuzzolo},
  {Olberg}, {Olbrich}, {Orfei}, {Orleanski}, {Ossenkopf}, {Peacock}, {Pearson},
  {Peron}, {Phillip-May}, {Piazzo}, {Planesas}, {Rataj}, {Ravera}, {Risacher},
  {Salez}, {Samoska}, {Saraceno}, {Schieder}, {Schlecht}, {Schl{\"o}der},
  {Schm{\"u}lling}, {Schultz}, {Schuster}, {Siebertz}, {Smit}, {Szczerba},
  {Shipman}, {Steinmetz}, {Stern}, {Stokroos}, {Teipen}, {Teyssier}, {Tils},
  {Trappe}, {van Baaren}, {van Leeuwen}, {van de Stadt}, {Visser}, {Wildeman},
  {Wafelbakker}, {Ward}, {Wesselius}, {Wild}, {Wulff}, {Wunsch}, {Tielens},
  {Zaal}, {Zirath}, {Zmuidzinas}, \& {Zwart}}]{degraauw2010}
{De Graauw}, T., {Helmich}, F.~P., {Phillips}, T.~G., {et~al.} 2010, \aap, 518,
  L6

\bibitem[{{Downes} \& {Eckart}(2007)}]{downes2007}
{Downes}, D. \& {Eckart}, A. 2007, \aap, 468, L57

\bibitem[{{Emprechtinger} {et~al.}(2013){Emprechtinger}, {Lis}, {Rolffs},
  {Schilke}, {Monje}, {Comito}, {Ceccarelli}, {Neufeld}, \& {van der
  Tak}}]{emprechtinger2013}
{Emprechtinger}, M., {Lis}, D.~C., {Rolffs}, R., {et~al.} 2013, \apj, 765, 61

\bibitem[{{Flagey} {et~al.}(2013){Flagey}, {Goldsmith}, {Lis}, {Gerin},
  {Neufeld}, {Sonnentrucker}, {De Luca}, {Godard}, {Goicoechea}, {Monje}, \&
  {Phillips}}]{flagey2013}
{Flagey}, N., {Goldsmith}, P.~F., {Lis}, D.~C., {et~al.} 2013, \apj, 762, 11

\bibitem[{{G{\'o}mez-Carrasco} {et~al.}(2014){G{\'o}mez-Carrasco}, {Godard},
  {Lique}, {Bulut}, {K{\l}os}, {Roncero}, {Aguado}, {Aoiz}, {Castillo},
  {Goicoechea}, {Etxaluze}, \& {Cernicharo}}]{gomezcarrasco2014}
{G{\'o}mez-Carrasco}, S., {Godard}, B., {Lique}, F., {et~al.} 2014, \apj, 794,
  33

\bibitem[{{Gonz{\'a}lez-Alfonso} {et~al.}(2014){Gonz{\'a}lez-Alfonso},
  {Fischer}, {Aalto}, \& {Falstad}}]{gonzalez2014}
{Gonz{\'a}lez-Alfonso}, E., {Fischer}, J., {Aalto}, S., \& {Falstad}, N. 2014,
  \aap, 567, A91

\bibitem[{{Gonz{\'a}lez-Alfonso} {et~al.}(2013){Gonz{\'a}lez-Alfonso},
  {Fischer}, {Bruderer}, {M{\"u}ller}, {Graci{\'a}-Carpio}, {Sturm}, {Lutz},
  {Poglitsch}, {Feuchtgruber}, {Veilleux}, {Contursi}, {Sternberg},
  {Hailey-Dunsheath}, {Verma}, {Christopher}, {Davies}, {Genzel}, \&
  {Tacconi}}]{gonzalez2013}
{Gonz{\'a}lez-Alfonso}, E., {Fischer}, J., {Bruderer}, S., {et~al.} 2013, \aap,
  550, A25

\bibitem[{{Gonz{\'a}lez-Alfonso} {et~al.}(2012){Gonz{\'a}lez-Alfonso},
  {Fischer}, {Graci{\'a}-Carpio}, {Sturm}, {Hailey-Dunsheath}, {Lutz},
  {Poglitsch}, {Contursi}, {Feuchtgruber}, {Veilleux}, {Spoon}, {Verma},
  {Christopher}, {Davies}, {Sternberg}, {Genzel}, \& {Tacconi}}]{gonzalez2012}
{Gonz{\'a}lez-Alfonso}, E., {Fischer}, J., {Graci{\'a}-Carpio}, J., {et~al.}
  2012, \aap, 541, A4

\bibitem[{{Goto} {et~al.}(2014){Goto}, {Geballe}, {Indriolo}, {Yusef-Zadeh},
  {Usuda}, {Henning}, \& {Oka}}]{goto2014}
{Goto}, M., {Geballe}, T.~R., {Indriolo}, N., {et~al.} 2014, \apj, 786, 96

\bibitem[{{Grenier} {et~al.}(2015){Grenier}, {Black}, \&
  {Strong}}]{grenier2015}
{Grenier}, I.~A., {Black}, J.~H., \& {Strong}, A.~W. 2015, \araa, 53, 199

\bibitem[{{Hollenbach} {et~al.}(2012){Hollenbach}, {Kaufman}, {Neufeld},
  {Wolfire}, \& {Goicoechea}}]{hollenbach2012}
{Hollenbach}, D., {Kaufman}, M.~J., {Neufeld}, D., {Wolfire}, M., \&
  {Goicoechea}, J.~R. 2012, \apj, 754, 105

\bibitem[{{Indriolo} \& {McCall}(2012)}]{mccall2012}
{Indriolo}, N. \& {McCall}, B.~J. 2012, \apj, 745, 91

\bibitem[{{Indriolo} {et~al.}(2012){Indriolo}, {Neufeld}, {Gerin}, {Geballe},
  {Black}, {Menten}, \& {Goicoechea}}]{indriolo2012}
{Indriolo}, N., {Neufeld}, D.~A., {Gerin}, M., {et~al.} 2012, \apj, 758, 83

\bibitem[{{Indriolo} {et~al.}(2015){Indriolo}, {Neufeld}, {Gerin}, {Schilke},
  {Benz}, {Winkel}, {Menten}, {Chambers}, {Black}, {Bruderer}, {Falgarone},
  {Godard}, {Goicoechea}, {Gupta}, {Lis}, {Ossenkopf}, {Persson},
  {Sonnentrucker}, {van der Tak}, {van Dishoeck}, {Wolfire}, \&
  {Wyrowski}}]{indriolo2015}
{Indriolo}, N., {Neufeld}, D.~A., {Gerin}, M., {et~al.} 2015, \apj, 800, 40

\bibitem[{{Israel} {et~al.}(2014){Israel}, {G{\"u}sten}, {Meijerink}, {Loenen},
  {Requena-Torres}, {Stutzki}, {van der Werf}, {Harris}, {Kramer},
  {Martin-Pintado}, \& {Weiss}}]{israel2014}
{Israel}, F.~P., {G{\"u}sten}, R., {Meijerink}, R., {et~al.} 2014, \aap, 562,
  A96

\bibitem[{{Kennicutt} \& {Evans}(2012)}]{kennicutt2012}
{Kennicutt}, R.~C. \& {Evans}, N.~J. 2012, \araa, 50, 531

\bibitem[{{Koribalski} {et~al.}(1995){Koribalski}, {Whiteoak}, \&
  {Houghton}}]{koribalski1995}
{Koribalski}, B., {Whiteoak}, J.~B., \& {Houghton}, S. 1995, \pasa, 12, 20

\bibitem[{{Kramer} {et~al.}(2005){Kramer}, {Mookerjea}, {Bayet},
  {Garcia-Burillo}, {Gerin}, {Israel}, {Stutzki}, \& {Wouterloot}}]{kramer2005}
{Kramer}, C., {Mookerjea}, B., {Bayet}, E., {et~al.} 2005, \aap, 441, 961

\bibitem[{{Lahuis} {et~al.}(2007){Lahuis}, {Spoon}, {Tielens}, {Doty}, {Armus},
  {Charmandaris}, {Houck}, {St{\"a}uber}, \& {van Dishoeck}}]{lahuis2007}
{Lahuis}, F., {Spoon}, H.~W.~W., {Tielens}, A.~G.~G.~M., {et~al.} 2007, \apj,
  659, 296

\bibitem[{{McElroy} {et~al.}(2013){McElroy}, {Walsh}, {Markwick}, {Cordiner},
  {Smith}, \& {Millar}}]{mcelroy2013}
{McElroy}, D., {Walsh}, C., {Markwick}, A.~J., {et~al.} 2013, \aap, 550, A36

\bibitem[{{Monje} {et~al.}(2014){Monje}, {Lord}, {Falgarone}, {Lis}, {Neufeld},
  {Phillips}, \& {G{\"u}sten}}]{monje2014}
{Monje}, R.~R., {Lord}, S., {Falgarone}, E., {et~al.} 2014, \apj, 785, 22

\bibitem[{{M{\"u}ller} {et~al.}(2005){M{\"u}ller}, {Schl{\"o}der}, {Stutzki},
  \& {Winnewisser}}]{mueller2005}
{M{\"u}ller}, H.~S.~P., {Schl{\"o}der}, F., {Stutzki}, J., \& {Winnewisser}, G.
  2005, Journal of Molecular Structure, 742, 215

\bibitem[{{Neufeld} {et~al.}(2010){Neufeld}, {Goicoechea}, {Sonnentrucker},
  {Black}, {Pearson}, {Yu}, {Phillips}, {Lis}, {de Luca}, {Herbst}, {Rimmer},
  {Gerin}, {Bell}, {Boulanger}, {Cernicharo}, {Coutens}, {Dartois},
  {Kazmierczak}, {Encrenaz}, {Falgarone}, {Geballe}, {Giesen}, {Godard},
  {Goldsmith}, {Gry}, {Gupta}, {Hennebelle}, {Hily-Blant}, {Joblin},
  {Ko{\l}os}, {Kre{\l}owski}, {Mart{\'{\i}}n-Pintado}, {Menten}, {Monje},
  {Mookerjea}, {Perault}, {Persson}, {Plume}, {Salez}, {Schlemmer}, {Schmidt},
  {Stutzki}, {Teyssier}, {Vastel}, {Cros}, {Klein}, {Lorenzani}, {Philipp},
  {Samoska}, {Shipman}, {Tielens}, {Szczerba}, \& {Zmuidzinas}}]{neufeld2010}
{Neufeld}, D.~A., {Goicoechea}, J.~R., {Sonnentrucker}, P., {et~al.} 2010,
  \aap, 521, L10

\bibitem[{{Oka} {et~al.}(2005){Oka}, {Geballe}, {Goto}, {Usuda}, \&
  {McCall}}]{oka2005}
{Oka}, T., {Geballe}, T.~R., {Goto}, M., {Usuda}, T., \& {McCall}, B.~J. 2005,
  \apj, 632, 882

\bibitem[{{Ott} {et~al.}(2001){Ott}, {Whiteoak}, {Henkel}, \&
  {Wielebinski}}]{ott2001}
{Ott}, M., {Whiteoak}, J.~B., {Henkel}, C., \& {Wielebinski}, R. 2001, \aap,
  372, 463

\bibitem[{{Ott}(2010)}]{ott2010}
{Ott}, S. 2010, in Astronomical Society of the Pacific Conference Series, Vol.
  434, Astronomical Data Analysis Software and Systems XIX, ed. Y.~{Mizumoto},
  K.-I. {Morita}, \& M.~{Ohishi}, 139

\bibitem[{{Papadopoulos}(2010)}]{papadopoulos2010}
{Papadopoulos}, P.~P. 2010, \apj, 720, 226

\bibitem[{{Persic} \& {Rephaeli}(2012)}]{persic2012}
{Persic}, M. \& {Rephaeli}, Y. 2012, Journal of Physics Conference Series, 355,
  012038

\bibitem[{{Pickett} {et~al.}(1998){Pickett}, {Poynter}, {Cohen}, {Delitsky},
  {Pearson}, \& {M{\"u}ller}}]{pickett1998}
{Pickett}, H.~M., {Poynter}, R.~L., {Cohen}, E.~A., {et~al.} 1998, \jqsrt, 60,
  883

\bibitem[{{Pilbratt} {et~al.}(2010){Pilbratt}, {Riedinger}, {Passvogel},
  {Crone}, {Doyle}, {Gageur}, {Heras}, {Jewell}, {Metcalfe}, {Ott}, \&
  {Schmidt}}]{pilbratt2010}
{Pilbratt}, G.~L., {Riedinger}, J.~R., {Passvogel}, T., {et~al.} 2010, \aap,
  518, L1

\bibitem[{{Porras} {et~al.}(2014){Porras}, {Federman}, {Welty}, \&
  {Ritchey}}]{porras2014}
{Porras}, A.~J., {Federman}, S.~R., {Welty}, D.~E., \& {Ritchey}, A.~M. 2014,
  \apjl, 781, L8

\bibitem[{{Roelfsema} {et~al.}(2012){Roelfsema}, {Helmich}, {Teyssier},
  {Ossenkopf}, {Morris}, {Olberg}, {Shipman}, {Risacher}, {Akyilmaz},
  {Assendorp}, {Avruch}, {Beintema}, {Biver}, {Boogert}, {Borys}, {Braine},
  {Caris}, {Caux}, {Cernicharo}, {Coeur-Joly}, {Comito}, {de Lange},
  {Delforge}, {Dieleman}, {Dubbeldam}, {de Graauw}, {Edwards}, {Fich},
  {Flederus}, {Gal}, {di Giorgio}, {Herpin}, {Higgins}, {Hoac}, {Huisman},
  {Jarchow}, {Jellema}, {de Jonge}, {Kester}, {Klein}, {Kooi}, {Kramer},
  {Laauwen}, {Larsson}, {Leinz}, {Lord}, {Lorenzani}, {Luinge}, {Marston},
  {Mart{\'{\i}}n-Pintado}, {McCoey}, {Melchior}, {Michalska}, {Moreno},
  {M{\"u}ller}, {Nowosielski}, {Okada}, {Orlea{\'n}ski}, {Phillips}, {Pearson},
  {Rabois}, {Ravera}, {Rector}, {Rengel}, {Sagawa}, {Salomons},
  {S{\'a}nchez-Su{\'a}rez}, {Schieder}, {Schl{\"o}der}, {Schm{\"u}lling},
  {Soldati}, {Stutzki}, {Thomas}, {Tielens}, {Vastel}, {Wildeman}, {Xie},
  {Xilouris}, {Wafelbakker}, {Whyborn}, {Zaal}, {Bell}, {Bjerkeli}, {De Beck},
  {Cavali{\'e}}, {Crockett}, {Hily-Blant}, {Kama}, {Kaminski}, {Lefl{\'o}ch},
  {Lombaert}, {de Luca}, {Makai}, {Marseille}, {Nagy}, {Pacheco}, {van der
  Wiel}, {Wang}, \& {Y{\i}ld{\i}z}}]{roelfsema2012}
{Roelfsema}, P.~R., {Helmich}, F.~P., {Teyssier}, D., {et~al.} 2012, \aap, 537,
  A17

\bibitem[{{Schilke} {et~al.}(2013){Schilke}, {Lis}, {Bergin}, {Higgins}, \&
  {Comito}}]{schilke2013}
{Schilke}, P., {Lis}, D.~C., {Bergin}, E.~A., {Higgins}, R., \& {Comito}, C.
  2013, Journal of Physical Chemistry A, 117, 9766

\bibitem[{{Sch{\"o}ier} {et~al.}(2005){Sch{\"o}ier}, {van der Tak}, {van
  Dishoeck}, \& {Black}}]{schoeier2005}
{Sch{\"o}ier}, F.~L., {van der Tak}, F.~F.~S., {van Dishoeck}, E.~F., \&
  {Black}, J.~H. 2005, \aap, 432, 369

\bibitem[{{Sofia} {et~al.}(2004){Sofia}, {Lauroesch}, {Meyer}, \&
  {Cartledge}}]{sofia2004}
{Sofia}, U.~J., {Lauroesch}, J.~T., {Meyer}, D.~M., \& {Cartledge}, S.~I.~B.
  2004, \apj, 605, 272

\bibitem[{{Sofia} {et~al.}(2011){Sofia}, {Parvathi}, {Babu}, \&
  {Murthy}}]{sofia2011}
{Sofia}, U.~J., {Parvathi}, V.~S., {Babu}, B.~R.~S., \& {Murthy}, J. 2011, \aj,
  141, 22

\bibitem[{{Sturm} {et~al.}(2011){Sturm}, {Gonz{\'a}lez-Alfonso}, {Veilleux},
  {Fischer}, {Graci{\'a}-Carpio}, {Hailey-Dunsheath}, {Contursi}, {Poglitsch},
  {Sternberg}, {Davies}, {Genzel}, {Lutz}, {Tacconi}, {Verma}, {Maiolino}, \&
  {de Jong}}]{sturm2011}
{Sturm}, E., {Gonz{\'a}lez-Alfonso}, E., {Veilleux}, S., {et~al.} 2011, \apjl,
  733, L16

\bibitem[{{Van der Hulst} {et~al.}(1983){Van der Hulst}, {Golisch}, \&
  {Haschick}}]{vdhulst1983}
{Van der Hulst}, J.~M., {Golisch}, W.~F., \& {Haschick}, A.~D. 1983, \apjl,
  264, L37

\bibitem[{{Van der Tak} {et~al.}(2008){Van der Tak}, {Aalto}, \&
  {Meijerink}}]{vdtak2008}
{Van der Tak}, F.~F.~S., {Aalto}, S., \& {Meijerink}, R. 2008, \aap, 477, L5

\bibitem[{{Van der Tak} {et~al.}(2006){Van der Tak}, {Belloche}, {Schilke},
  {G{\"u}sten}, {Philipp}, {Comito}, {Bergman}, \& {Nyman}}]{vdtak2006}
{Van der Tak}, F.~F.~S., {Belloche}, A., {Schilke}, P., {et~al.} 2006, \aap,
  454, L99

\bibitem[{{Van der Tak} {et~al.}(2007){Van der Tak}, {Black}, {Sch{\"o}ier},
  {Jansen}, \& {van Dishoeck}}]{vdtak2007}
{Van der Tak}, F.~F.~S., {Black}, J.~H., {Sch{\"o}ier}, F.~L., {Jansen}, D.~J.,
  \& {van Dishoeck}, E.~F. 2007, \aap, 468, 627

\bibitem[{{Van der Tak} {et~al.}(2013{\natexlab{a}}){Van der Tak},
  {Chavarr{\'{\i}}a}, {Herpin}, {Wyrowski}, {Walmsley}, {van Dishoeck}, {Benz},
  {Bergin}, {Caselli}, {Hogerheijde}, {Johnstone}, {Kristensen}, {Liseau},
  {Nisini}, \& {Tafalla}}]{vdtak2013wish}
{Van der Tak}, F.~F.~S., {Chavarr{\'{\i}}a}, L., {Herpin}, F., {et~al.}
  2013{\natexlab{a}}, \aap, 554, A83

\bibitem[{{Van der Tak} {et~al.}(2013{\natexlab{b}}){Van der Tak}, {Nagy},
  {Ossenkopf}, {Makai}, {Black}, {Faure}, {Gerin}, \& {Bergin}}]{vdtak2013oh+}
{Van der Tak}, F.~F.~S., {Nagy}, Z., {Ossenkopf}, V., {et~al.}
  2013{\natexlab{b}}, \aap, 560, A95

\bibitem[{{Van der Tak} \& {van Dishoeck}(2000)}]{vdtak2000}
{Van der Tak}, F.~F.~S. \& {van Dishoeck}, E.~F. 2000, \aap, 358, L79

\bibitem[{{Van Dishoeck} {et~al.}(2013){Van Dishoeck}, {Herbst}, \&
  {Neufeld}}]{vandishoeck2013}
{Van Dishoeck}, E.~F., {Herbst}, E., \& {Neufeld}, D.~A. 2013, Chemical
  Reviews, 113, 9043

\bibitem[{{Walter} {et~al.}(2002){Walter}, {Weiss}, \& {Scoville}}]{walter2002}
{Walter}, F., {Weiss}, A., \& {Scoville}, N. 2002, \apjl, 580, L21

\bibitem[{{Wang} {et~al.}(2004){Wang}, {Henkel}, {Chin}, {Whiteoak}, {Hunt
  Cunningham}, {Mauersberger}, \& {Muders}}]{wang2004}
{Wang}, M., {Henkel}, C., {Chin}, Y.-N., {et~al.} 2004, \aap, 422, 883

\bibitem[{{Wei{\ss}} {et~al.}(2008){Wei{\ss}}, {Kov{\'a}cs}, {G{\"u}sten},
  {Menten}, {Schuller}, {Siringo}, \& {Kreysa}}]{weiss2008}
{Wei{\ss}}, A., {Kov{\'a}cs}, A., {G{\"u}sten}, R., {et~al.} 2008, \aap, 490,
  77

\bibitem[{{Wei{\ss}} {et~al.}(2010){Wei{\ss}}, {Requena-Torres}, {G{\"u}sten},
  {Garc{\'{\i}}a-Burillo}, {Harris}, {Israel}, {Klein}, {Kramer}, {Lord},
  {Martin-Pintado}, {R{\"o}llig}, {Stutzki}, {Szczerba}, {van der Werf},
  {Philipp-May}, {Yorke}, {Akyilmaz}, {Gal}, {Higgins}, {Marston}, {Roberts},
  {Schl{\"o}der}, {Schultz}, {Teyssier}, {Whyborn}, \& {Wunsch}}]{weiss2010}
{Wei{\ss}}, A., {Requena-Torres}, M.~A., {G{\"u}sten}, R., {et~al.} 2010, \aap,
  521, L1

\bibitem[{{Wyrowski} {et~al.}(2010{\natexlab{a}}){Wyrowski}, {Menten},
  {G{\"u}sten}, \& {Belloche}}]{wyrowski2010oh+}
{Wyrowski}, F., {Menten}, K.~M., {G{\"u}sten}, R., \& {Belloche}, A.
  2010{\natexlab{a}}, \aap, 518, A26

\bibitem[{{Wyrowski} {et~al.}(2010{\natexlab{b}}){Wyrowski}, {van der Tak},
  {Herpin}, {Baudry}, {Bontemps}, {Chavarria}, {Frieswijk}, {Jacq},
  {Marseille}, {Shipman}, {van Dishoeck}, {Benz}, {Caselli}, {Hogerheijde},
  {Johnstone}, {Liseau}, {Bachiller}, {Benedettini}, {Bergin}, {Bjerkeli},
  {Blake}, {Braine}, {Bruderer}, {Cernicharo}, {Codella}, {Daniel}, {di
  Giorgio}, {Dominik}, {Doty}, {Encrenaz}, {Fich}, {Fuente}, {Giannini},
  {Goicoechea}, {de Graauw}, {Helmich}, {Herczeg}, {J{\o}rgensen},
  {Kristensen}, {Larsson}, {Lis}, {McCoey}, {Melnick}, {Nisini}, {Olberg},
  {Parise}, {Pearson}, {Plume}, {Risacher}, {Santiago}, {Saraceno}, {Tafalla},
  {van Kempen}, {Visser}, {Wampfler}, {Y{\i}ld{\i}z}, {Black}, {Falgarone},
  {Gerin}, {Roelfsema}, {Dieleman}, {Beintema}, {de Jonge}, {Whyborn},
  {Stutzki}, \& {Ossenkopf}}]{wyrowski2010h2o+}
{Wyrowski}, F., {van der Tak}, F., {Herpin}, F., {et~al.} 2010{\natexlab{b}},
  \aap, 521, L34

\bibitem[{{Yang} {et~al.}(2013){Yang}, {Gao}, {Omont}, {Liu}, {Isaak},
  {Downes}, {van der Werf}, \& {Lu}}]{yang2013}
{Yang}, C., {Gao}, Y., {Omont}, A., {et~al.} 2013, \apjl, 771, L24

\bibitem[{{Yun} {et~al.}(1993){Yun}, {Ho}, \& {Lo}}]{yun1993}
{Yun}, M.~S., {Ho}, P.~T.~P., \& {Lo}, K.~Y. 1993, \apjl, 411, L17

\bibitem[{{Zhao} {et~al.}(2015){Zhao}, {Galazutdinov}, {Linnartz}, \&
  {Kre{\l}owski}}]{zhao2015}
{Zhao}, D., {Galazutdinov}, G.~A., {Linnartz}, H., \& {Kre{\l}owski}, J. 2015,
  \apjl, 805, L12

\end{thebibliography}

\end{document}